\documentstyle[prl,aps,preprint]{revtex}

\begin{document}

\draft
\title{Gravitino Warm Dark Matter Motivated by the CDF $ee\gamma\gamma $
Event}
\author{M. Kawasaki}
\address{Institute for Cosmic Ray Research, The University of Tokyo,
  Tanashi, Tokyo 188, Japan}
\author{Naoshi Sugiyama}
\address{Department of Physics, 
Kyoto University,
Kyoto 606-01, Japan}
\author{T. Yanagida}
\address{Department of Physics, School of Science, The University of
  Tokyo, Tokyo 113, Japan}
\date{\today}

\maketitle

\begin{abstract}
    The $ee\gamma\gamma +\rlap/E_T$ event observed by the CDF at
    Fermilab is naturally explained by dynamically supersymmetry
    breaking models and suggests the presence of the light gravitino
    which can be a warm dark matter. We consider large scale structure
    of the universe in the worm dark matter model and find that the
    warm dark matter plays almost the same role in the formation of
    the large scale structure as a cold dark matter if its mass
    is about $0.5$keV.  We also study the Ly~$\alpha$ absorption
    systems which are presumed to be galaxies at high redshifts and
    show that the baryon density in the damped Ly~$\alpha$ absorption
    systems predicted by the warm dark matter model is quite
    consistent with the present observation.
\end{abstract}

\pacs{14.80, 95.35, 98.54.A, 98.65.-r}


Low-energy supersymmetry(SUSY) is a very attractive candidate beyond
the standard model, since it provides a natural solution to the gauge
hierarchy problem~\cite{Maiani,Veltman}. If there exists the SUSY it
must be spontaneously broken. The hidden sector model in $N=1$
supergravity~\cite{Nilles} is widely used for a realization of the
SUSY breaking. Although this model has many attractive features, it
suffers from a serious cosmological problem, i.e. the Polonyi
problem~\cite{Coughlan}. There is, however, no such a problem in an
alternative model~\cite{Dine} where the SUSY is broken dynamically by
some new strong gauge interactions.  In this class of models the SUSY
breaking is mediated to the ordinary sector by the ordinary gauge
interactions and another problem in the SUSY standard model, i.e. the
flavor changing neutral current problem, is also solved automatically.

This dynamical model predicts the SUSY breaking scale $F$ to be low as
$(100 - 1000)$TeV and the gravitino mass $m_{3/2}$ in the range of 10eV
$-$ 1keV. Therefore, the usual lightest SUSY particle (LSP)
decays into the gravitino \footnote{
The longitudinal component of the gravitino (Goldstino) couples to
matter with strength proportional to $F^{-1} \sim
1/\sqrt{m_{3/2}M_{p}}$ ($M_p$: Planck mass). Thus the decay is very
fast for the light gravitino.  }
and cannot be a stable cold dark matter (CDM) in the universe. Instead
of it the gravitino is a true LSP and can form a dark matter.  Because
its mass is so small, the gravitino has larger a velocity dispersion
than the CDM.  Such a type of dark matter is called a warm dark matter
(WDM)~\cite{Warm}.

It has been, recently, pointed out~\cite{Dimopoulos} that the
dynamical SUSY-breaking model naturally explains the $ee\gamma\gamma +
\rlap/E_T$ event observed by the CDF experiment~\cite{CDF}.  The event
is explained by sequent decays~\cite{Dimopoulos};
$\tilde{e}^{-}(\tilde{e}^{+}) \rightarrow e^{-}(e^{+}) + \tilde{B}$
and $\tilde{B} \rightarrow \gamma + \tilde{G}$ where $\tilde{e}$,
$\tilde{B}$ and $\tilde{G}$ are selectron, bino and gravitino,
respectively. The decay length of the bino into a photon is given by
\begin{equation}
    \label{decay-length}
    c\tau_{\tilde{B}} \simeq 5 
    \left(\frac{M_{\tilde{B}}}{100 {\rm GeV}}\right)^{-5}
    \left(\frac{m_{3/2}}{0.5{\rm keV}}\right)^{2}{\rm m},
\end{equation}
where $M_{\tilde{B}}$ is the bino mass which should be $(38 - 100)$GeV
~\cite{Dimopoulos}. For $m_{3/2} \lesssim 0.5$keV and
$M_{\tilde{B}}=100$GeV, the decay length is less than about 5m,
which is consistent with the bino decay inside the CDF detector.
Thus, the light gravitino is well motivated.

In this letter, we show that the warm (gravitino) dark matter whose
mass is about 0.5keV~\footnote{
If we take, e.g. $m_{3/2} \simeq 0.3$keV, we get $c\tau_{\tilde{B}}
\simeq 2$m. In this case the density parameter of the bino is $\simeq
0.5$ and hence we need an additional contribution to $\Omega$ to get
the flat universe. However the results in the text are unchanged if
the additional contribution comes from CDM~\cite{DGP}.}
plays almost the same role in the large-scale
structure formation as CDM.  We also study damped Ly~$\alpha$
absorption systems which are presumed to be the progenitors of
present-day spiral galaxies in the WDM model and find that the
observed mass of neutral hydrogens in Ly~$\alpha$ systems is
consistent with the prediction by the WDM model.  Colombi {\it et
al.}~\cite{CDW} has recently studied the large scale structure
formations by WDM.  However they have considered a very light WDM ($
\sim 100$eV).  Such a light WDM may cause a serious problem against
the damped Ly~$\alpha$ systems.

Before the universe becomes colder than the gravitino mass ($T \gtrsim
10^7 (m_{\rm 3/2}/1{\rm keV})$K), the gravitino behaves as a
relativistic particle.  Therefore the free streaming of the gravitino
smears out small-scale density fluctuations and leads to a sharp
cutoff in the power spectrum of the density fluctuations. The cutoff
scale ( $=$ free streaming scale) is given by
\begin{equation}
    \label{cutoff}
    R_{\rm fs} = 0.2\left(\frac{g}{100}\right)^{-4/3}(\Omega h^2)^{-1} 
    {\rm Mpc},
\end{equation}
where $g$ is the effective number of particle degrees of freedom when
the gravitino decoupled ($g\simeq 200$ for the particle content of the
minimal SUSY standard model and hereafter we take $g=200$), $h$ the
Hubble constant in units of 100km/s/Mpc and $\Omega$ the present
density parameter of the gravitino which is related to the gravitino
mass $m_{3/2}$ by $\Omega h^2 = (g/100)^{-1} (m_{3/2}/{\rm keV})$ .
Since we only consider a gravitino dominated universe, $\Omega \simeq
\Omega_0$ where $\Omega_0$ is the total density parameter at present.
Assuming a scale invariant Harrison-Zeldovich spectrum, we can write
the power spectrum $P(k)$ of WDM as~\cite{Bardeen}
\begin{eqnarray}
    P(k) & = & Ak|T(k)|^2, \\
    T(k) & = & \exp\left[ -\frac{kR_{\rm fs}}{2}
        - \frac{(kR_{\rm fs})^2}{2}\right] T_0(k),\\
    T_0 & = &\frac{\ln(1+2.34q)}{2.34q} 
         [1+3.89q  \\
     & & +(16.1q)^2+(5.46q)^3+(6.71q)^4]^{-1/4},
\end{eqnarray}
where $q$ is defined as $q \equiv k/\Omega_0
h^2/\exp(-\Omega_B-\sqrt{h/0.5}\Omega_B/\Omega_0) {\rm Mpc}^{-1}$
taking into account the dependence on the baryon density
$\Omega_B$~\cite{Sugi}, $T(k)$ is the transfer functions for WDM, and
$A$ the normalization constant which is determined by COBE DMR 4 year
data~\cite{COBE}.  Notice that for a CDM-dominated universe the power
spectrum $P_{\rm CDM}$ is given by $P_{\rm CDM}(k) = Ak|T_0(k)|^2$.
The WDM power spectrum for $\Omega_0=1, h=0.5$ are shown in Fig.~1
together with the CDM power spectrum with the same cosmological
parameters.  Here, we have taken $m_{3/2} = 0.5$keV corresponding to
the $\Omega_0=1$ universe. Since the cutoff scale is relatively small
($\sim$ 0.3Mpc) the power spectrum relevant for the large-scale
structure ($k\lesssim 1h {\rm Mpc}^{-1}$) is almost the same as the
CDM one. (This contrasts the hot dark matter spectrum which has a
cutoff of the order of 0.1Mpc$^{-1}$.)  Therefore, the WDM model with
$\Omega_0=1$ has the same problem as CDM one, i.e.  the shape and the
magnitude do not fit the observational data from the galaxy
surveys~\cite{PD} which are also shown in Fig.~1. In particular, the
amplitude of the power spectrum normalized by COBE is too large for
$k= (0.03 - 0.3)h$Mpc$^{-1}$.  The amplitude contradicts not only the
galaxy surveys but also the recent analysis of velocity
fields~\cite{Zar} (shaded region) which could directly reflect the
mass distribution.  If we take smaller value for $A$, the power
spectrum $P(k)$ better fits to the data. Since the tensor
mode~\cite{tensor} or isocurvature mode~\cite{KSY} may significantly
contribute to $\delta T/T$ in COBE scales, it is possible that the
actual normalization of $A$ is smaller. For example, we show the power
spectrum normalized by $\sigma_8 = 0.8$ with the same cosmological
parameters in Fig.~1. Here $\sigma_8$ is the mass overdensity within
spheres of radius $8h^{-1}$Mpc.  Notice that the COBE normalization
gives $\sigma_8 = 1.2$. As is seen in Fig.~1 $P(k)$ with $\sigma_8 =
0.8$ is in a good agreement with the velocity field data.

The difference between WDM and CDM is more significant for galaxies or
smaller systems. The existence of the cutoff in the WDM spectrum
delays the galaxy formation compared with in the CDM case. Damped
Ly~$\alpha$ absorption systems observed in QSO spectra are important
since they give us information about galactic systems in the early
universe. It is presumed that the damped Ly~$\alpha$ absorptions
observed in QSO spectra are due to neutral hydrogens contained in
galactic systems at high redshifts ($z \sim 1 - 4$). Therefore, the
observed damped Ly~$\alpha$ absorptions can give us interesting
information about baryons contained in the galactic
systems~\cite{Wolfe} and can set a constraint on galaxy formation
models. In fact, this constraint is very stringent for the mixed
dark matter (MDM $=$ hot $+$ cold dark matter) model~\cite{Mo,Ma}
since few galaxies are formed at high redshifts in the MDM model.

Here we study the damped Ly~$\alpha$ constraint on the WDM model.
Following ref.~\cite{Mo}, we use the Press-Schechter
theory~\cite{Press} to estimate the comoving number density $N(z,M)$
of the dark matter halos with mass between $M$ and $M+dM$ at redshift
$z$:
\begin{eqnarray}
    \label{Press}
    N(z,M)dM & = & \sqrt{\frac{2}{\pi}}
    \frac{\rho_0}{M}\frac{\delta_c}{D_1(z)}
    \left[-\frac{1}{\sigma^2(M)}\frac{\partial \sigma (M)}{\partial M}
        \right] \nonumber \\
    & & \times
    \exp\left[-\frac{\delta_c^2}{2\sigma^2(M)D_1^2(z)}\right],
\end{eqnarray}
where $\rho_0$ is the mean comoving mass density, $\delta_c$ is the
overdensity threshold for the collapse ( $= 1.68$, corresponding to
the prediction of the spherical collapse model), $D_1(z)$ is the
function for the growth of the perturbations ($D_1(z) = (1+z)^{-1}$
for $\Omega_0 =1$) and $\sigma^2(M)$ is the rms mass fluctuation in a
top-hat window with radius $r_M= [M/(4\pi\rho_0/3)]^{1/3}$, given by
\begin{eqnarray}
    \sigma^2(M)& = & \frac{1}{2\pi^2}\int P(k)W^2(kr_M)k^2dk,\\
    W(x) & = & \frac{3}{x^3}(\sin x - x \cos x). \nonumber
\end{eqnarray}
Since we assume that the damped Ly~$\alpha$ absorptions are due to the
neutral hydrogens in galactic systems, we need identify the halos with
a certain mass range as galaxies. For this purpose, it is
convenient to use circular velocity $v_c$ which is related to $M$ by
\begin{eqnarray}
    M & = & 2.45\times 10^{11}M_{\odot}(1+z)^{-3/2}
    (\Omega_0 h^2)^{-1/2} \nonumber \\
    \label{velocity}
    & & \times (v_c/100{\rm km/s})^3 \Omega_0^{0.3}.
\end{eqnarray}
For the spiral galaxies the circular velocity $v_c$ is in the range of
$100 - 250$km/s. Smaller object may also contribute to the absorption.
However the lower limit may not be less than $50$km/s since baryons in
such small halo cannot be cooled enough to form a gaseous
disk~\cite{Efstathiou}. The upper limit is also uncertain since
gaseous disks might survive when halo merging occurs~\cite{Mo}. In
Fig.~2(a) we show the density parameter of baryons in galactic systems
($\Omega_D$) predicted in the WDM model for $\Omega_0 =1, h=0.5$, and
$\Omega_B=0.05$ together with observational data~\cite{Wolfe}.  Notice
that the adopted $\Omega_B$ is consistent with the prediction by big
bang nucleosynthesis ($\Omega_Bh^2 = 0.0125 \pm
0.0025$~\cite{Walker}).  In this case, the predicted $\Omega_D$ with
$v_c = 100 - 250$km/s is above the data. Since some fraction of
baryons become stars and do not contribute to the absorption, the data
should be taken to be lower limit to $\Omega_D$. Therefore the
prediction by the WDM model is quite consistent with the observation.
For comparison, we also show the predictions by CDM and MDM in
Fig.~2(a). The MDM model ($\Omega_{\rm hot} \simeq 0.3$ and
$\Omega_{\rm cold} \simeq 0.7$) can explain the data of the large
scale structure in the universe better than the CDM
model~\cite{Klypin}.  However, as already mentioned, since the galaxy
formation in the MDM model seems too late, it is difficult to explain
the damped Ly~$\alpha$ absorption~\cite{Mo,Ma}. One can also consider
the (warm $+$ hot) dark matter model~\cite{Borgani} which may explain
the large scale structure of the universe.  However, this model has
the same difficulty in explaining the damped Ly~$\alpha$ absorption
systems as the MDM model. In the case of CDM, the predicted $\Omega_D$
is larger than that predicted by WDM at higher redshifts. Thus, the
survey of the damped Ly~$\alpha$ systems at $z \gtrsim 5$ may possibly
distinguish two models.

As is seen before, the WDM power spectrum with $\Omega_0 =1$ and COBE
normalization gives poor fit to the data at large scales, which leads
us to consider the WDM power spectrum with smaller normalization
constant. In Fig.~2(b) the predicted $\Omega_D$ is shown for $\Omega_0
=1, h=0.5, \Omega_B=0.05, \sigma_8 = 0.8$. In the figure, it is seen
that the WDM model is still quite consistent with the observation even if
we take the smaller normalization for the power spectrum.

If the light fermion forms dark matter in a galactic halo, its phase
space density in the galactic core might be larger than that allowed
by the Fermi statistics~\cite{Tremaine}.  This phase space constraint
puts a stringent constraint on the mass of the dark matter fermion.
From the study on the stellar motions in dwarf galaxies, the mass
should be larger than about 500eV. Since the analysis of dwarf
galaxies may contain systematic errors, this constraint may not be
taken seriously. In our case, the dark matter is gravitino and its
mass is about 500eV for $\Omega_0 = 1$ and $h=0.5$. Therefore the phase
space constraint is satisfied even if we take the constraint from the
dwarf galaxies.

In summary, the $ee\gamma\gamma + \rlap/E_T$ event observed by CDF is
naturally explained by dynamical SUSY-breaking models and suggests the
light gravitino. The light gravitino is a LSP and can be a WDM. We
study the formation of the large scale structure of the universe and
the adamped Ly~$\alpha$ absorption systems in the WDM model. It is
found that the WDM plays almost the same role in the formation of the
large scale structure as the CDM if its mass is about $0.5$keV. The
difference between CDM and WDM becomes more significant when one
considers the damped Ly~$\alpha$ absorption systems which are presumed
to be galaxies at high redshifts.  The baryon density in the damped
Ly~$\alpha$ absorption systems predicted by the WDM model is quite
consistent with the observational data. The future observation of
damped Ly~$\alpha$ systems at higher redshifts may distinguish the WDM
model from the CDM one.  Measurements of cosmic microwave background
anisotropies may provide another possible way to distinguish WDM from
CDM~\cite{DS} although the difference is very small and only appears
on fine angular scales.  We need to wait for new generation satellite
experiments(e.g., MAP and COBRAS/SAMBA).

\begin{figure}
    \caption{ The matter power spectra $P(k)$ in WDM models 
    for the $\Omega_0=1, h=0.5$ and $\Omega_B=0.05$ with $n=1$
    adiabatic fluctuations. The power spectra normalized by the COBE 4
    year data(solid) and by $\sigma_8=0.8$ (dotted) are plotted.  An
    adiabatic CDM model with COBE 4 year normalization is also plotted
    (dashed).  The observational data of galaxy surveys are taken from
    Peacock and Dodds~\protect\cite{PD}. Shaded regions are the best
    fitted value of Mark III catalog of peculiar velocities of
    galaxies by Zaroubi et al.~\protect\cite{Zar} with $30\%$ errors.}
\end{figure}

\begin{figure}
    \caption{ (a)Evolution of the Density parameter $\Omega_D$ of the
    baryons contained in galactic systems whose circular velocity is
    between $100$ and $250$km/s using Press-Schechter theory and the
    power spectrum normalized by COBE.  Solid, dashed and
    dotted-dashed curves denote the predictions by WDM, CDM and MDM,
    respectively. Symbols represent for the observational data by
    Wolfe {\it et al.\/}~\protect\cite{Wolfe}.(b) Same as (a)  with
    the power spectrum normalized by $\sigma_8 = 0.8$.}
\end{figure}

\end{document}